\documentclass[11pt]{article}
\usepackage{moriond}
\usepackage{amsmath}
\usepackage{graphicx}
\usepackage{amssymb}
\usepackage{color}
\usepackage{soul}

\def\be{\begin{equation}}
\def\ee{\end{equation}}
\def\bea{\begin{eqnarray}}
\def\eea{\end{eqnarray}}

\def\zmin{z_{\rm min}}
\def\zmax{z_{\rm max}}
\def\met{\not{\!\!\rm E}_T}

\newcommand{\Photo}{}

\begin{document}
ANL-HEP-CP-13-27, IIT-CAPP-13-01
\vspace*{4cm}
\title{MEASURING TOP-QUARK POLARIZATION IN TOP-PAIR + MISSING ENERGY EVENTS}

\author{Edmond L. Berger$^{b}$, Qing-Hong Cao$^{a}$, Jiang-Hao Yu$^{c}$, Hao Zhang$^{b,d}$}
\address{$^a$Department of Physics and State Key Laboratory of Nuclear Physics
and Technology, Peking University, Beijing 100871, China\\$^b$High Energy Physics Division, Argonne National Laboratory, Argonne, IL 60439, USA\\$^c$Theory Group, Department of Physics, The University 
of Texas at Austin, Austin, TX 78712, USA\\$^d$Illinois Institute of Technology, Chicago, IL 60616-3793, USA}

\maketitle\abstracts{
The polarization of a top-quark can be sensitive to new physics beyond the standard model. 
We propose a novel method to measure top-quark polarization,
based on the charged lepton energy fraction in top-quark decay, and illustrate the method 
with a detailed simulation 
of top-quark pairs produced in supersymmetric top squark pair production.  We show that the 
lepton energy ratio distribution that we define is very sensitive to the top-quark polarization but 
insensitive to the precise measurement of the top-quark energy.}

\section{Introduction}
Events with a top-quark pair plus missing energy ($t\bar{t}+\!\met$) are promising channels in which to investigate models of new physics (NP) beyond the standard model (SM).  Missing energy originates typically from non-interacting or otherwise invisible dark matter (DM) candidates in the NP models, along with neutrinos from SM weak decays.  In these events the polarization of the top quark is sensitive to the chirality structure of the top quark's interaction with a postulated parent new heavy resonance and the DM.  The top-quark polarization might provide a new way to gain insight into NP models.  Measurements of the top-quark polarization tend to rely on the predicted angular correlation of the momentum of a charged lepton (from the top-quark decay) with the top-quark spin~\cite{Kane:1991bg}.  However, this measurement is difficult in $t\bar{t}+\met$ events because it is generally not possible to reconstruct the top-quark kinematics, i.e., to  disentangle the kinematic effects of the DM particles from neutrinos that accompany the charged leptons in the top-quark leptonic decay.  

We define and examine the energy fraction of the charged lepton from the top quark as a novel measure of top-quark polarization,  without the requirement of top-quark reconstruction and knowledge of the dark matter mass and spin~\cite{Berger:2012an}.  
We emphasize a few advantages of our energy ratio variable:  (i) it is sensitive to the top-quark polarization; (ii) it is not sensitive to the mass splitting  between a heavy resonance parent and the DM candidate, provided that this splitting is not too small; (iii) the difference between the left-handed ($t_L$) and the right-handed top quark ($t_R$) is not sensitive to the spin of a heavy parent resonance or to the collider energy.

\section{The method} 
In the leptonic decay of a top quark, $t\to b W^+ \to b \ell^+ \nu$, the 
correlation of the momentum of the charged lepton $\ell^+$ with the polarization $\hat{s}_t$ 
of the top-quark, viewed in the top-quark rest frame, takes the form 
$(1+\hat{s}_t z)/2$, where $z \equiv \cos\theta$ is the cosine of the angle between 
the top-quark spin axis and the lepton momentum.   
\begin{figure}[t]
\includegraphics[scale=0.4,clip]{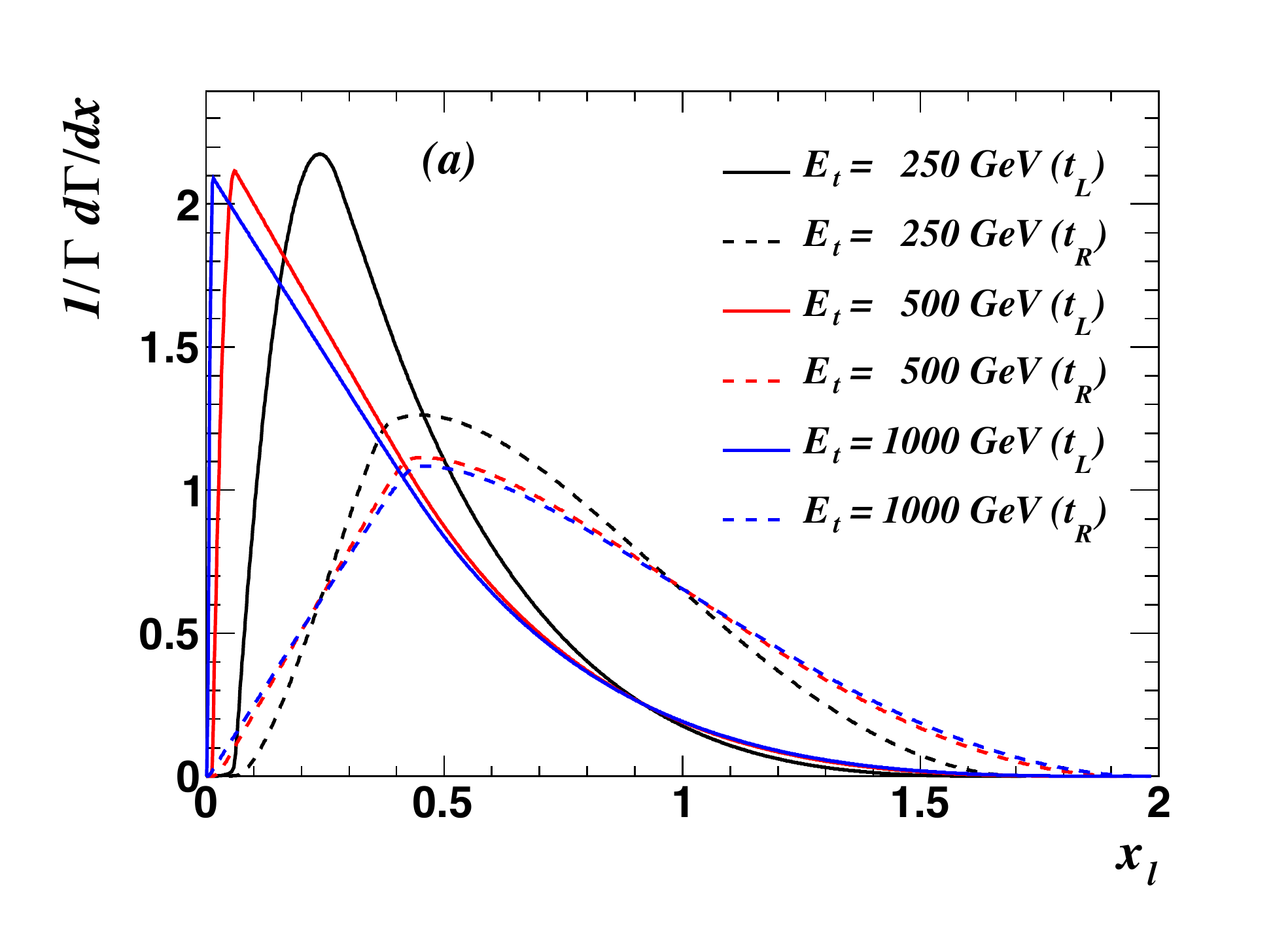}
\includegraphics[scale=0.4,clip]{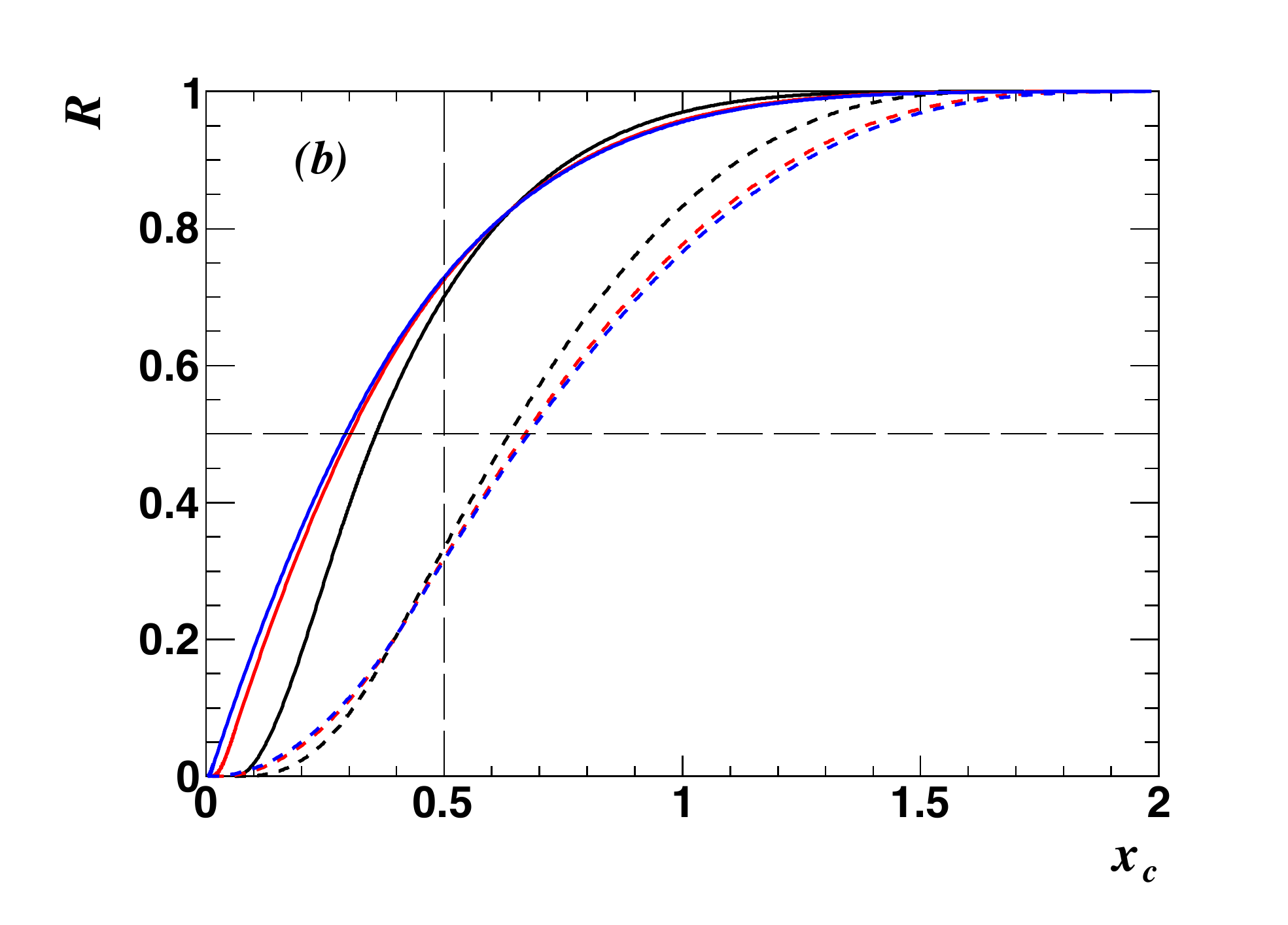}
\caption{
(a)Distributions of the energy fraction $x_{\ell}$ of a charged lepton from top-quark decay for $E_{t}=250,~500,~1000~{\rm GeV}$;  (b) The ratio 
$\mathcal R$ as a function of the cut threshold $x_c$ for $E_{t}=250,~500,~1000~{\rm GeV}$. 
The solid lines represent left-handed top-quark decay while the dashed lines represent right-handed top quarks.  
\label{fig:en_lep} }
\end{figure}
For a boosted top quark with energy $E_t$, the distribution in the energy fraction $x_{\ell} \equiv 2 E_\ell / E_t$ of the charged lepton becomes   
\be
\frac{d\Gamma(\hat{s}_t )}{dx}=\frac{\alpha_{W}^{2}m_{t}}{64\pi AB}\int_{\zmin}^{\zmax}
x\gamma^{2}[1-x\gamma^{2}(1-z\beta)]
\left(1+\hat{s}_{t}\frac{z-\beta}{1-z\beta}\right)
\text{Arctan}\biggl[\frac{Ax\gamma^{2}(1-z\beta)}{B-x\gamma^{2}(1-z\beta)}\biggr]dz. 
\label{eq:energy}
\ee
Here, $A=\Gamma_{W}/m_{W}$ is the ratio of the $W$-boson width and $W$-boson mass, $B=m_{W}^{2}/m_{t}^{2}\approx 0.216$ is the squared ratio of the $W$-boson mass and top-quark mass, and the limits of integration are $\zmin = {\rm max}[(1-1/\gamma^2 x)/\beta,-1]$, $\zmax = {\rm min}[(1-B/\gamma^2 x)/\beta,1]$ with $\gamma = E_t/m_t$ and $\beta = \sqrt{1-1/\gamma^{2}}$. 
The function $\text{Arctan}$ is defined as 
$\arctan(x)$ for $x\ge 0$ while $\pi+\arctan(x)$ for $x<0$.
We introduce a ratio $\mathcal{R}$ as a quantitative measure of the energy fractions of $t_L$ and $t_R$, 
\be
\mathcal{R}(x_c)=\frac{\displaystyle 1}{\displaystyle \Gamma} \int_{0}^{x_c}\frac{d\Gamma}{d x_\ell} dx_\ell
\equiv \frac{\Gamma(x_\ell<x_c)}{\Gamma}.  
\ee
This ratio is a function of the the cut threshold $x_c$ of the energy fraction $x_\ell$. 
An analytic expression can be derived for $\mathcal{R}(x_c)$ in the limit $\beta \to 1$.  It takes the form 
\be
\mathcal{R}(x_c)=\frac{3x_c(1-\lambda_t)}
{2(1+2B)}-\frac{3\lambda_t x_c^2(1-B+\ln B)}{2(1+2B)(1-B)^2},  
\label{eq:anal1}
\ee
for $x_c \in (0,~2B)$, and 
\be
\mathcal{R}(x_c)=\frac{B^2(2B-3)}{(1+2B)(1-B)^2}+\frac{3x_c(1-\lambda_t)}{2(1-B)^2(1+2B)}-\frac{3x_c^2[1+2\lambda_t\ln(x_c/2)]}{4(1-B)^2(1+2B)}+\frac{x_c^3(1+3\lambda_t)}{8(1-B)^2(1+2B)}
\label{eq:anal2}
\ee
for $x_c\in (2B,~2)$, where $\lambda_t=(-1, +1)$ for $(t_L, t_R)$, respectively.  For small $x_c$, these expressions show that $\mathcal{R}(x_c)$ grows linearly with $x_c$ for $t_L$, whereas $\mathcal{R}(x_c)$ grows as  $x_c^2$ for $t_R$ ($\lambda_t =1$).    

The analytic expressions Eqs.~\ref{eq:anal1} and~\ref{eq:anal2} also explain why the curves for $E_t=500~{\rm GeV}$ ($\beta=0.94$) and $E_t=1000~{\rm GeV}$ ($\beta=0.99$) almost overlap.  For an energetic top quark, an important consequence is that the difference between $\mathcal{R}(x_c)$ for $t_L$ and $t_R$ is not sensitive to $E_t$, i. e., the mass splitting between the parent heavy resonance and the DM candidate, as long as the mass splitting is not too small.  The $t_L$ and $t_R$ curves are insensitive to the origin of the top quark in the collision, whether from a heavy fermion decay or from a scalar decay.  In other words, $\mathcal{R}(x_c)$ quantifies the top-quark polarization but not the top-quark origin.   Moreover, in order to extract NP signal events from SM backgrounds, one must normally impose a set of hard kinematic cuts on the leptons and jets in the final state.   These hard cuts force the top quark to be very energetic and thus to satisfy the limit $\beta \to 1$.  Therefore, another virtue of the $\mathcal{R}(x_c)$ variable is that the difference between the
$t_L$ and $t_R$ curves does not vary with the hard cuts. 
To prove this method useful, we turn to an explicit calculation of top squark ($\tilde{t}$) pair production, $pp \to \tilde{t} \overline{\tilde{t}} X \to t \bar{t} \tilde{\chi} \tilde{\chi} X$.  

\section{Collider simulation}
We perform a parton-level Monte Carlo simulation of top squark ($\tilde{t}$) pair production 
$pp \to \tilde{t} \overline{\tilde{t}} X \to t \bar{t} \tilde{\chi} \tilde{\chi} X$ to demonstrate that $\mathcal{R}$ remains useful for distinguishing $t_L$ and $t_R$ even when $E_t$ cannot be measured directly.   We assume the colored scalar $\tilde{t}$ decays entirely into 
$t\tilde{\chi}$ through the effective coupling $\mathcal{L}_{\tilde{t}t\tilde{\chi}}= g_{\rm eff} \tilde{t} \tilde{\chi}(\cos\theta_{\rm eff} P_L + \sin\theta_{\rm eff} P_R) t$, where the angle $\theta_{\rm eff}$ depends on the mass matrix mixings of the top squark and the neutralino sector, and 
$P_{\rm L/R}$ is the usual left/right-handed projector.   
Our benchmark point has $m_{\tilde{t}}=350~{\rm GeV}$ and a representative DM mass $m_{\tilde{\chi}}=50~{\rm GeV}$.   
We simulate $\tilde{t}$ pair production with decay to a top-quark pair plus DM candidates at the LHC with 8 TeV energy.   We demand that the top-quark decays semi-leptonically, and that the antitop-quark decays hadronically.   
Two irreducible SM backgrounds, $t\bar{t}$ and $t\bar{t}Z$ production, are considered.
The details of the simulation can be found in our paper~\cite{Berger:2012an}.
After all cuts, we find that the numbers of signal and background 
events are 130 and 22 at 8 TeV and 20~fb$^{-1}$ integrated luminosity, 
for a signal significance of $S/\sqrt B = 28$.  

In $\tilde{t}$ pair production the decay chains of $\tilde{t} \to t \tilde{\chi}$ and  $\tilde{\bar{t}} \to \bar{t} \tilde{\chi}$ have similar 
kinematics because the heavy $\tilde{t}$'s are not highly boosted.
In this work we investigate the energy of the antitop quark as an estimator of the top-quark energy, with the antitop-quark 
required to decay into three jets~\footnote{Another useful variable in the literature is $E_\ell/(E_\ell + E_b)$ where the $b$-jet and $\ell^+$ originate from the same top-quark decay~\cite{Shelton:2008nq}. }.  
We define a new energy fraction variable $x^\prime_{\ell}= 2 E_{\ell} / E_{\bar{t}}$.
After convolution with the production cross section, a ratio $\mathcal{R}^\prime$ can be defined as 
\be
\mathcal{R}^\prime(x_c^\prime)=\frac{1}{\sigma({\rm tot})}{\displaystyle \int_{0}^{x_c^\prime}
\frac{d\sigma}{d x^\prime_\ell} dx^\prime_\ell} 
\equiv \frac{\sigma(x^\prime_\ell<x_c^\prime)}{\sigma({\rm tot})},
\ee
where $d\sigma/d x_\ell^\prime$ is the differential cross section, and $x^\prime_c$ is the cut threshold of $x^\prime_\ell$. 

We use a $\chi^2$-template method based on the $W$ boson and 
top-quark masses to select the three jets from the hadronic decay of the antitop quark.  
For each event we pick the combination which minimizes $\chi^2=(m_W-m_{jj})^2/
\Delta m_W^2  +(m_{t}-m_{jjj})^2/\Delta m_t^2$, where $\Delta m_W$ and 
$\Delta m_t$ are the width of the $W$-boson and the top-quark, respectively.
The efficiency of this method is 84\%.
\begin{figure}[t]
\includegraphics[scale=0.4,clip]{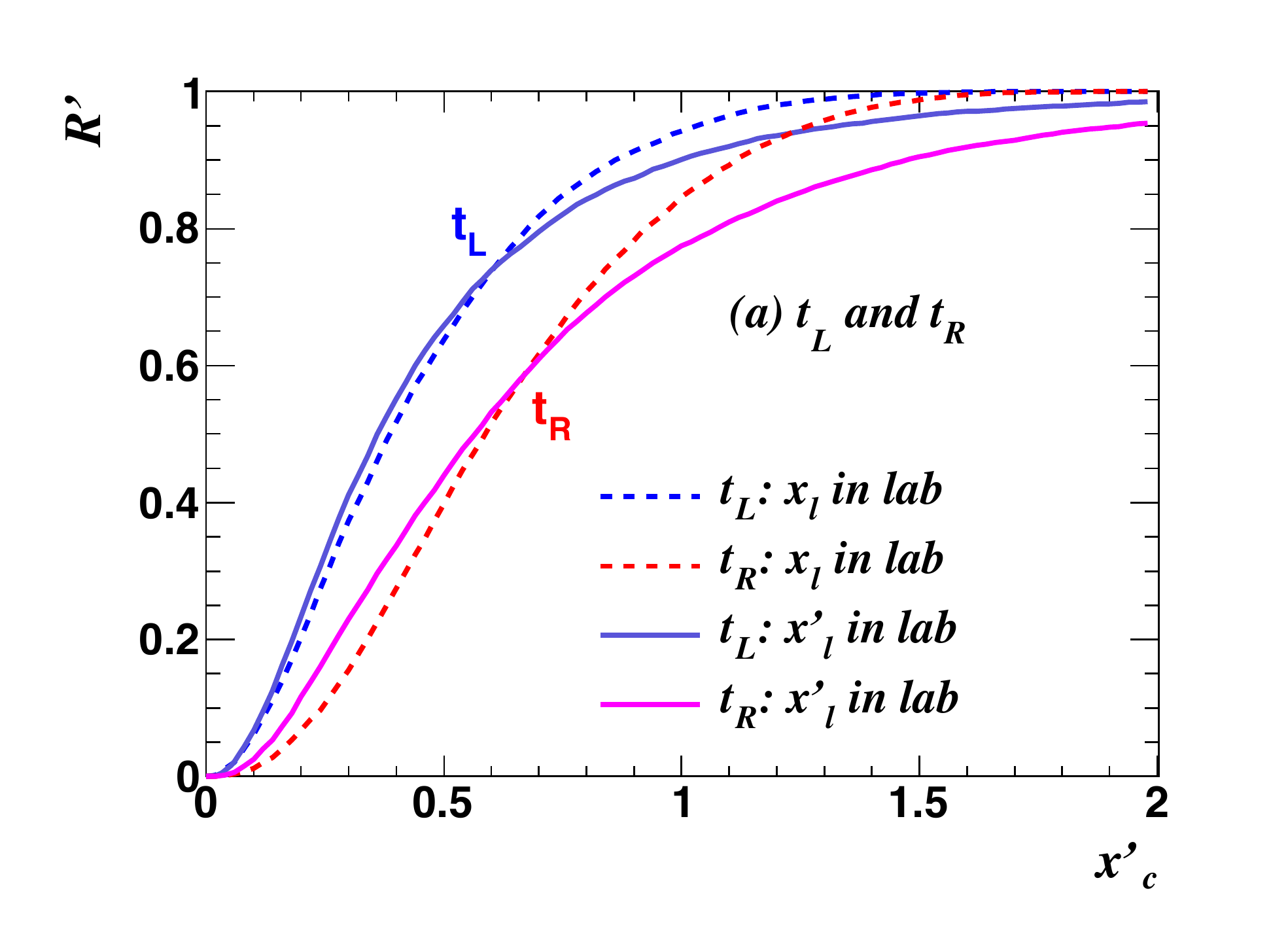}
\includegraphics[scale=0.4,clip]{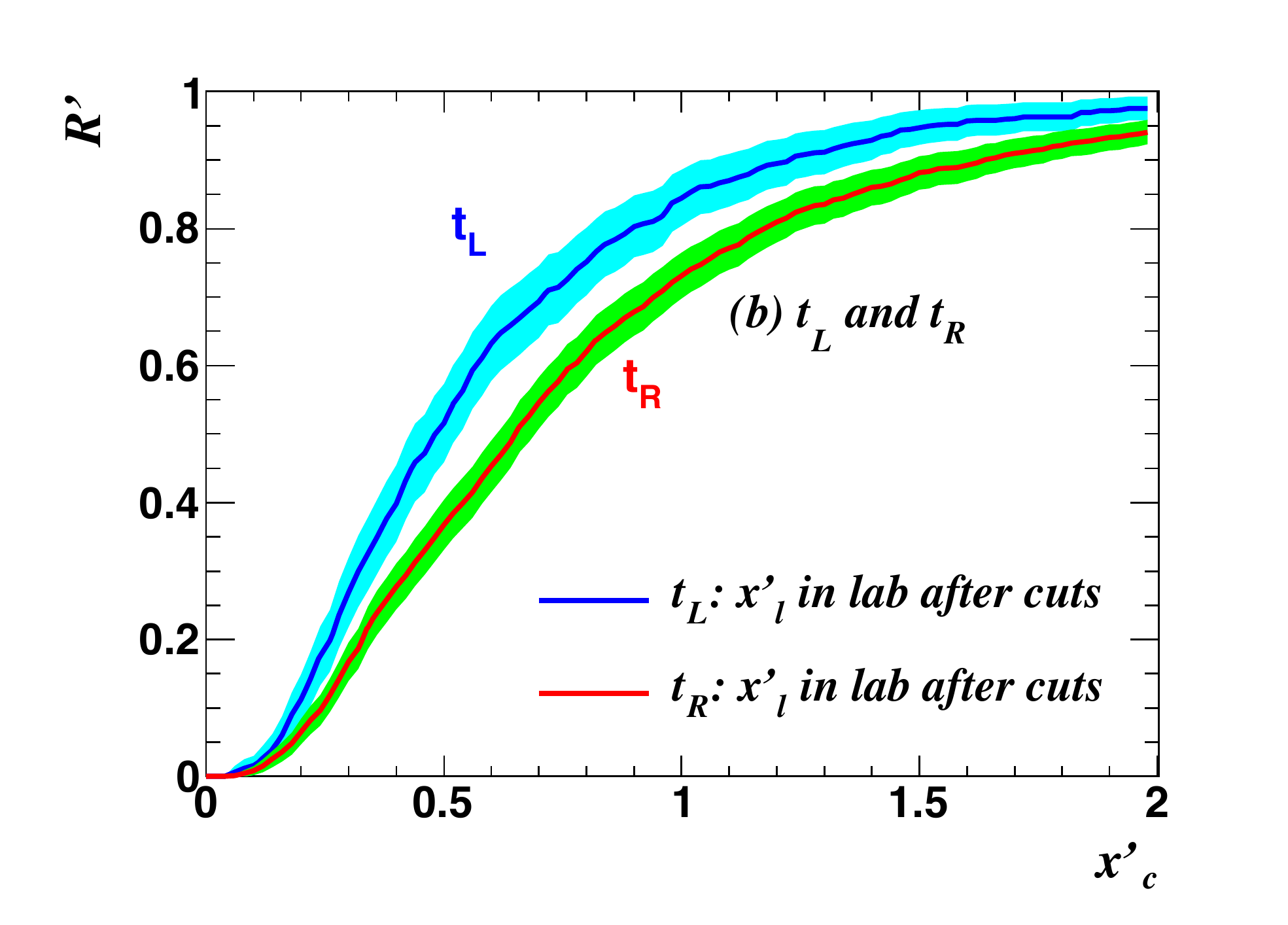}
\caption{(a) The $\mathcal R$ distributions as a function of the cut threshold $x^\prime_c$ for a 350~GeV $\tilde{t}$ quark with 
pure right-handed or left-handed couplings  
at the LHC with 8 TeV energy. The lepton energy fraction is evaluated in the lab frame from either $\bar{t}_{\rm had}^{\rm rec}$ 
($x'_{\ell}$) or the top-quark energy $t_{\rm lep}$ ($x_{\ell}$). With 
$\sin\theta_{\rm eff}=1$ ($\cos\theta_{\rm eff}=1$) the top quark is mainly right-handed (left-handed), and the curves are labeled by $t_R$ 
($t_L$). 
(b) The statistical uncertainty band of $\mathcal R$ is shown for both $t_L$ and $t_R$ at the $1\sigma$ confidence level 
for an assumed $20$ fb$^{-1}$ of integrated luminosity.   
\label{fig:mssm_cut} }
\end{figure}
After the antitop-quark energy is reconstructed in the lab frame, $\mathcal R^\prime$ can be obtained with its cut threshold $x^\prime_c$ dependence.

Armed with both the Monte Carlo level momenta and the reconstructed momenta, we perform several comparisons to evaluate how faithful the $\mathcal R^\prime$ distribution is to the true $\mathcal R$~
(Fig.~\ref{fig:mssm_cut}).  
We conclude that $x_c^\prime$ is a good variable when $x_c$ cannot be obtained. 
Lastly, comparing $\mathcal{R}$ at the Monte Carlo level and at the reconstruction level, we see a slight downward shift for both $t_L$ and 
$t_R$.   This effect arises because the $p_T$ cuts on the lepton reduce the number of events with $x^\prime_{\ell} < x^\prime_c$.
  
The results in Fig.~\ref{fig:mssm_cut} (a) establish that $x_c^\prime$ is a suitable variable and that $\mathcal R^\prime$ serves as a good substitute for $\mathcal R$.  After all the cuts, about 87 (101) 
 signal events are needed to distinguish $t_L$ ($t_R$) from an unpolarized top quark at 95\% confidence level (C.L.) in the absence of background, with only about 23 signal events to discriminate $t_R$ from $t_L$. Although the benchmark point chosen has been subsequently excluded~\cite{LHC:2013stop}, 
the method proposed in this work remains powerful for masses in the allowed region.
  
Thus far in this section, we assume the $t$-$\tilde{t}$-$\tilde{\chi}$ coupling is completely left-handed or right-handed, but in general the coupling is a mixture of both.  Once data are obtained, we could use the $\mathcal{R}^\prime$ curves shown in Fig.~\ref{fig:mssm_cut}(b) 
as templates in fits to these data to extract $\theta_{\rm eff}$ and shed light on the nature of top squark mixing.  

\section{Other implications}
Our method can be applied to several NP models.  We performed a detailed simulation of pair production of a $T$-odd top-quark partner ($T_-$) in
the Littlest Higgs Model with T-parity (LHT), $pp\to T_-\overline{T}_- X \to t\bar{t} A_H A_H X$,  
where $A_H$ is the $T$-odd photon partner.     
Our numerical results are very similar to those shown for $t_R$ in Fig.~\ref{fig:mssm_cut}.    
Verification of mainly right-handed polarization would provide a powerful check of the model~\cite{Cao:2006wk}.  Another example is the leptophobic $Z^\prime$ boson, which couples only to the SM quarks. 
The top-quark polarization could be used to probe the handedness of the $Z^\prime$-$q$-$q$ coupling which is sensitive to how the SM quarks are gauged under the new gauge symmetry~\cite{Berger:2011hn}.

\section*{Acknowledgments}
The work by E.L.B. and H.Z. is supported in part by the U.S.
DOE under Grant No.~DE-AC02-06CH11357. H.Z. is also supported by DOE under the Grant No.~DE-FG02-94ER40840. 
Q.H.C. is supported by the National Natural Science Foundation of China under Grant No. 11245003.
J.H.Y. is supported by the U.S. National Science Foundation 
under Grant No. PHY-0855561.

\end{document}